\documentclass{emulateapj}

\usepackage{float}
\usepackage{amsmath}
\usepackage{epsfig,floatflt}

\usepackage{graphicx}
\usepackage[]{subfigure}
\usepackage{color}
\definecolor{Blue}{rgb}{0.1,0.1,1.0} 
\definecolor{Magenta}{rgb}{1.0,0.1,0.5} 
\definecolor{LRed}{rgb}{0.8,0.0,0.0}

\newcommand{\nc}{\newcommand}

\nc{\be}[1]{\begin{equation}\mbox{$\label{#1}$}}
\nc{\bea}[1]{\begin{eqnarray} \mbox{$\label{#1}$}}
\nc{\Section}[2]{\section{#2}\label{#1}}
\nc{\Bibitem}[1]{\bibitem{#1}}
\nc{\Label}[1]{\label{#1}}

\nc{\eea}{\end{eqnarray}}
\nc{\ee}{\end{equation}}

\nc{\bdm}{\begin{displaymath}}
\nc{\edm}{\end{displaymath}}
\nc{\dpsty}{\displaystyle}
\nc{\bc}{\begin{center}}
\nc{\ec}{\end{center}}
\nc{\ea}{\end{array}}
\nc{\bab}{\begin{abstract}}
\nc{\eab}{\end{abstract}}
\nc{\btab}{\begin{tabular}}
\nc{\etab}{\end{tabular}}
\nc{\bit}{\begin{itemize}}
\nc{\eit}{\end{itemize}}
\nc{\ben}{\begin{enumerate}}
\nc{\een}{\end{enumerate}}
\nc{\bfig}{\begin{figure}}
\nc{\efig}{\end{figure}}

\nc{\arreq}{&\!=\!&}
\nc{\arrmi}{&\!-\!&}
\nc{\arrpl}{&\!+\!&}
\nc{\arrap}{&\!\!\!\approx\!\!\!&}
\nc{\non}{\nonumber}

\def\lsim{\; \raise0.3ex\hbox{$<$\kern-0.75em
      \raise-1.1ex\hbox{$\sim$}}\; }
\def\gsim{\; \raise0.3ex\hbox{$>$\kern-0.75em
      \raise-1.1ex\hbox{$\sim$}}\; }

\nc{\DOT}{\hspace{-0.08in}{\bf .}\hspace{0.1in}}
\nc{\Laada}{\hbox {$\sqcap$ \kern -1em $\sqcup$}}
\nc\loota{{\scriptstyle\sqcap\kern-0.55em\hbox{$\scriptstyle\sqcup$}}}
\nc\Loota{{\sqcap\kern-0.65em\hbox{$\sqcup$}}}
\nc\laada{\Loota}
\nc{\qed}{\hskip 3em \hbox{\BOX} \vskip 2ex}

\nc{\real}{{\rm I \! R}}
\nc{\Z}{{\sf Z \!\!\! Z}}
\nc{\complex}{{\rm C\!\!\! {\sf I}\,\,}}
\def\bigid{\leavevmode\hbox{\small1\kern-3.8pt\normalsize1}}
\def\id{\leavevmode\hbox{\small1\kern-3.3pt\normalsize1}}
\nc{\slask}{\!\!\!/}
\nc{\bis}{{\prime\prime}}
\nc{\pa}{\partial}
\nc{\ra}{\rangle}
\nc{\goto}{\rightarrow}
\nc{\swap}{\leftrightarrow}

\nc{\EE}[1]{ \mbox{$\cdot10^{#1}$} }
\nc{\abs}[1]{\left|#1\right|}
\nc{\at}[2]{\left.#1\right|_{#2}}
\nc{\norm}[1]{\|#1\|}
\nc{\abscut}[2]{\Abs{#1}_{\scriptscriptstyle#2}}
\nc{\vek}[1]{{\rm\bf #1}}
\nc{\integral}[2]{\int\limits_{#1}^{#2}}
\nc{\inv}[1]{\frac{1}{#1}}
\nc{\dd}[2]{{{\partial #1}\over{\partial #2}}}
\nc{\ddd}[2]{{{{\partial}^2 #1}\over{\partial {#2}^2}}}
\nc{\dddd}[3]{{{{\partial}^2 #1}\over
    {\partial #2 \partial #3}}}
\nc{\dder}[2]{{{d #1}\over{d #2}}}
\nc{\ddder}[2]{{{d^2 #1}\over{d {#2}^2}}}
\nc{\dddder}[3]{{d^2 #1}\over
    {d #2 d #3}}
\nc{\dx}[1]{d\,^{#1}x}
\nc{\dy}[1]{d\,^{#1}y}
\nc{\dz}[1]{d\,^{#1}z}
\nc{\dl}[1]{\frac{d\,^{#1}l}{(2\pi)^{#1}}}
\nc{\dk}[1]{\frac{d\,^{#1}k}{(2\pi)^{#1}}}
\nc{\dq}[1]{\frac{d\,^{#1}q}{(2\pi)^{#1}}}

\nc{\bfT}{{\bf T }}

\nc{\cA}{{\cal A}}
\nc{\cB}{{\cal B}}
\nc{\cD}{{\cal D}}
\nc{\cE}{{\cal E}}
\nc{\cG}{{\cal G}}
\nc{\cH}{{\cal H}}
\nc{\cL}{{\cal L}}
\nc{\cO}{{\cal O}}
\nc{\cT}{{\cal T}}
\nc{\cN}{{\cal N}}
\nc{\cR}{{\cal R}}

\nc{\rvac}[1]{|{\cal O}#1\rangle}
\nc{\lvac}[1]{\langle{\cal O}#1|}
\nc{\rvacb}[1]{|{\cal O}_\beta #1\rangle}
\nc{\lvacb}[1]{\langle{\cal O}_\beta #1 |}
\nc{\bb}{\bar{\beta}}
\nc{\bt}{\tilde{\beta}}
\nc{\ctH}{\tilde{\cal H}}
\nc{\chH}{\hat{\cal H}}
\nc{\al}{\alpha}
\nc{\g}{\gamma}
\nc{\Del}{\Delta}
\nc{\e}{\textrm{e}}
\nc{\eps}{\epsilon}
\nc{\lam}{\lambda}
\nc{\Om}{\Omega}
\nc{\ve}{\varepsilon}
\nc{\mn}{{\mu\nu}}
\nc{\vp}{\varphi}

\nc{\rf}[1]{(\ref{#1})}
\nc{\nn}{\nonumber \\*}
\nc{\bfB}{\bf{B}}
\nc{\bfv}{\bf{v}}
\nc{\bfx}{\bf{x}}
\nc{\bfy}{\bf{y}}
\nc{\vx}{\vec{x}}
\nc{\vy}{\vec{y}}
\nc{\oB}{\overline{B}}
\nc{\oI}{\overline{I}}
\nc{\oR}{\overline{R}}
\nc{\rar}{\rightarrow}
\nc{\ti}{\times}
\nc{\slsh}{\hskip-5pt/}
\nc{\sm}{Standard~Model~}
\nc{\MP}{M_{\rm Pl}}
\nc{\mpl}{M_{\rm Pl}}
\nc{\tp}{t_{\rm Pl}}

\nc{\pmin}{p_{\rm min}}
\nc{\pmax}{p_{\rm max}}
\nc{\fo}{f_0}
\nc{\foi}{f_{0,i}\,}
\nc{\fop}{f_0^P}
\nc{\fou}{f_0^U}

\nc{\eff}{{\rm eff}}
\nc{\MT}{M_{\rm T}}
\nc{\ML}{M_{\rm L}}
\nc{\kk}{\vek{k}}
\nc{\pp}{{\rm p}}
\nc{\pt}{\partial_t}
\nc{\half}{{1\over 2}}
\nc{\w}{\omega}
\nc{\uhat}{\hat{U}_\w}

\nc{\etal}{\mbox{\it et al.}}
\nc{\ie}{{\it i.e. }}
\nc{\eg}{{\it e.g. }}
\nc{\trh}{T_{\rm RH}}
\nc{\ad}{{a'\over a}}
\nc{\bd}{{b'\over b}}
\nc{\Rd}{{R'\over R}}
\nc{\diag}{{\textrm{diag}}}
\nc{\mato}[1]{\tilde{#1}}
\nc{\sinn}{\textrm{sinn}}
\nc{\sech}{\textrm{sech}}
\nc{\I}{\textrm{I}}
\nc{\II}{\textrm{II}}
\nc{\III}{\textrm{III}}
\nc{\vev}[1]{\langle #1 \rangle}
\nc{\hyp}{\,\; F_{1{\hskip -16pt}2}{\hskip 11pt}}
\nc{\brhom}{\overline{\rho}_M}
\nc{\brho}{\overline{\rho}}
\nc{\rhob}{\overline{\rho}}
\nc{\Pb}{\overline{P}}
\nc{\bH}{\overline{H}}
\nc{\ep}{{1+4\eps}}

\nc{\deriv}[2]{ 
\frac{\mathrm{d}#1}{\mathrm{d}#2}
}
\nc{\Mnu}{M_\nu}
\nc{\bee}{\begin{equation}}
\nc{\ene}{\end{equation}}
\nc{\hdp}{\sigma_8 (\Omega_{\rm m}/0.3)^{0.37}}
\nc{\avis}{\alpha_{vis}}
\nc{\cvis}{c^2_{vis}}
\nc{\clam}{c^2_{lam}}

\def\smiley{\hbox{\large$\bigcirc$\hspace{-.80em}%
\raise.2ex\hbox{$\cdot\cdot$}\kern-.61em    
\lower.2ex\hbox{\scriptsize$\smile$}}\ }

\def\frowney{\hbox{\large$\bigcirc$\hspace{-.80em}%
\raise.2ex\hbox{$\cdot\cdot$}\kern-.635em
\lower.2ex\hbox{\scriptsize$\frown$}}\ }

\begin{document}

\title{Bayesian analysis of white noise levels in the 5-year WMAP data}
\author{N. E. Groeneboom\altaffilmark{1}, H. K. Eriksen,
  \altaffilmark{1}, K. Gorski \altaffilmark{2}, G. Huey
  \altaffilmark{2}, J. Jewell \altaffilmark{2}, B. Wandelt \altaffilmark{3}}

\email{nicolaag@astro.uio.no}

\altaffiltext{1}{Institute of Theoretical Astrophysics, University of
  Oslo, P.O.\ Box 1029 Blindern, N-0315 Oslo, Norway}

\altaffiltext{2}{Jet Propulsion Laboratory, California Institute of Technology, Pasadena CA91109}

\altaffiltext{3}{Departments of Physics and Astronomy, 
University of Illinois at Urbana-Champaign, 
1002W. Green Street, Urbana, IL61801 }

\date{\today}

\begin{abstract} 
  We develop a new Bayesian method for estimating white noise levels
  in CMB sky maps, and apply this algorithm to the 5-year WMAP data.
  We assume that the amplitude of the noise RMS is scaled by a
  constant value, $\alpha$, relative to a pre-specified noise
  level. We then derive the corresponding conditional density,
  $P(\alpha \,| \, s, C_\ell, d)$, which is subsequently integrated
  into a general CMB Gibbs sampler. We first verify our code by
  analyzing simulated data sets, and then apply the framework to the
  WMAP data. For the foreground-reduced 5-year WMAP sky maps and
    the nominal noise levels initially provided in the 5-year data
    release, we find that the posterior means typically range between
  $\alpha=1.005\pm0.001$ and $\alpha=1.010\pm0.001$ depending on
  differencing assembly, indicating that the noise level of these maps
  are biased low by 0.5-1.0\%. The same problem is not observed
  for the uncorrected WMAP sky maps.  After the preprint version
    of this letter appeared 
    on astro-ph., the WMAP team has corrected the values presented on
    their web page, noting that the initially provided values were in
    fact estimates from the 3-year data release, not from the 5-year
    estimates. However, internally in their 5-year analysis the
    correct noise values were used, and no cosmological results are
    therefore compromised by this error. Thus, our method has already
    been demonstrated in practice to be both useful and accurate.
\end{abstract}

\keywords{cosmic microwave background --- cosmology: observations --- 
methods: numerical}

\maketitle

\section{Introduction}
\label{sec:introduction}

The cosmic microwave background (CMB) is probably the most valuable
source of  observational data in modern cosmology. Several
experiments have been carried out to map its anisotropies, most
notably the Wilkinson Microwave Anisotropy Map (WMAP)
\citep{bennett:2003, hinshaw:2007}. The WMAP experiment has provided
unique new insights in the workings of the universe, from large to
small scales, and we now believe that we understand the
main physical process from after inflation and up until today.

The theory of inflation was initially proposed as a solution to the
horizon and flatness problem \citep{guth:1981}. Additionally, it
established a highly successful theory for the formation of primordial
density perturbations, thus providing the required seeds for the
large-scale structures (LSS), later giving rise to the temperature
anisotropies in the cosmic microwave background radiation that we
observe today \citep{guth:1981,linde:1982, muhkanov:1981,
  starobinsky:1982, linde:1983, linde:1994, smoot:1992, ruhl:2003,
  runyan:2003, scott:2003}.

From this theory, we are able to predict what the statistical
properties of the CMB map should be, given a set of cosmological
parameters. The goal of the cosmological data analyst is then to
determine how well this universe model fits real-life data, which is
contaminated by foregrounds, systematics and various
uncertainties. The end result may be summarized in terms of a
joint posterior including all unknown quantities, from which the
desired cosmological parameters  may be obtained by marginalizing
over any relevant nuisance parameters.

A long-time discussion within the field of cosmological data analysis
has revolved around the process of optimally extracting
the underlying cosmological signal from the data. One general and
potentially optimal framework for doing this, based on a statistical
algorithm called Gibbs sampling, was first described by
\citet{jewell:2004}, \citet{wandelt:2004} and
\citet{eriksen:2004}, and extended and applied to the WMAP data
by \citet{odwyer:2004, eriksen:2007a, eriksen:2007b, larson:2007,
  eriksen:2008a,eriksen:2008b, groeneboom:2009}. This algorithm provides the user with
samples drawn from a joint CMB posterior, which can also include a
large number of external nuisance parameters. One well-known and
important example of this is joint component separation and CMB
power spectrum estimation.

We have developed an independent implementation of the CMB Gibbs
sampler which we call ``Slave'', corresponding to a lightweight C++
version of ``Commander'', described by \citet{eriksen:2004}. While
Slave may lack advanced features such as foreground estimation and
multi-band analysis, it does include all the basic features
required for elementary CMB power spectrum analysis (e.g., support
for cut-sky analysis, anisotropic noise distribution), and it also
takes advantage of the object oriented design features available in
C++. This is particularly useful when adding additional features into
the Gibbs sampler, as exemplified in the present paper.

In this paper, we apply this new implementation to a new practical
problem, and consider estimation of the white noise level of CMB sky
maps directly from the data. Specifically, we develop the necessary
machinery for including this operation into the CMB Gibbs sampler, and
apply this tool to the 5-year WMAP data.

\section{Methods}
\label{sec:methods}

An observed CMB map may be modeled as: 
\begin{equation}
 d = As + n
\end{equation}
where $A$ denotes the instrumental beam, $s$ is the desired CMB
signal, and $n$ is instrumental noise. The noise is in this
  paper assumed to be uncorrelated, and the corresponding noise
covariance matrix in pixel space is thus $N_{ij} = \sigma_i^2
\delta_{ij}$, where $\sigma_{i}$ is the noise standard deviation for
the $i$th pixel. Further, we assume that the CMB fluctuations are
Gaussian and isotropic, so the signal covariance matrix simplifies to
$C_{\ell m,\ell' m'} = C_\ell \delta_{\ell \ell'}\delta_{m m'}$. 

In order to estimate the power spectrum $C_{\ell}$ and the signal $s$
given the data, we need to sample from the joint distribution
$P(C_\ell, s | d)$. The algorithm for sampling from this distribution
 by Gibbs sampling is described exstensivly by
\citet{jewell:2004}, \citet{wandelt:2004}, citet{chu:2005} and \citet{eriksen:2004}.
A self-contained pedagogical introduction to the Gibbs sampling algorithm
is presented in \cite{groeneboom:2009b}, together with a presentation
of the ``Slave'' framework.

The standard Gibbs sampler draws samples from the joint distribution,
$P(s,C_{\ell}|d)$, by alternately sampling from the conditional
distributions $P(s|C_\ell, d)$ and $P(C_\ell | s)$. If one wishes to
introduce further parameters into the data model, all that is required
for joint estimation with the existing parameters is a sampling
algorithm for the corresponding conditional distribution.

\begin{figure}
\mbox{\epsfig{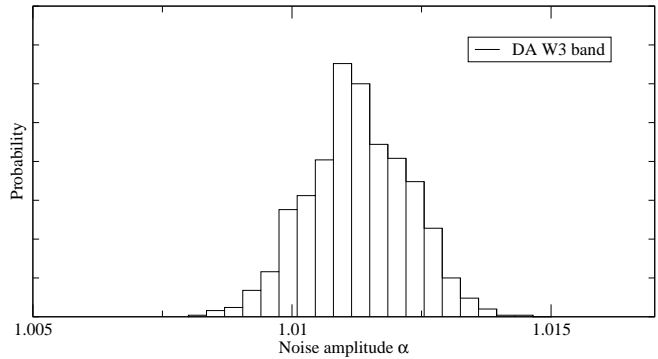}}
\caption{Posterior distribution $P(\alpha|d)$ for the
  foreground-corrected W3 channel. This channel exhibits the strongest
  detection of $\alpha \ne 1$ of any DA.}
\label{fig:w3}
\end{figure}

Traditionally, the noise properties used in the Gibbs sampler
\citep[e.g.,][]{eriksen:2004} have been assumed known to infinite
precision. In this paper, however, we relax this assumption, and
introduce a new free parameter, $\alpha$, that scales the fiducial
noise covariance matrix, $N^{\textrm{fid}}$,, such that $N = \alpha
N^{\textrm{fid}}$. Thus, if there is no deviation between the assumed and
real noise levels, then $\alpha$ should equal 1.

The full joint posterior, $P(s,C_\ell, \alpha \,|\, d)$, now includes
the amplitude $\alpha$. We can rewrite this as follows:
\begin{equation}
\label{eq:postnoise}
  P(s,C_\ell, \alpha \,|\,d) = P(d \,|\,s, \alpha) \cdot P(s, C_\ell) \cdot P(\alpha)
\end{equation}
where the first term is the likelihood,
\begin{equation}
  P(d \,| \,s, \alpha) = \frac{e^{-\frac{1}{2}(d-s)(\alpha N)^{-1}(d-s)
    }}{\sqrt{|\alpha N|}},
\end{equation}
the second term is a CMB prior, and the third term is a prior on
$\alpha$. Note that the latter two are independent, given that
these describe two a-priori independent objects.  In this paper, we
adopt a Gaussian prior centered on unity on $\alpha$, $P(\alpha) \sim
N(1,\sigma_\alpha^2)$. Typically, we choose a very loose prior,
such that the posterior is completely data-driven.

The conditional distribution for $\alpha$ can now be expressed as
\begin{equation}
  P(\alpha \,|\, s,C_\ell, d) \propto
  \frac{e^{-\frac{\beta}{2\alpha}}}{\alpha^{n/2}}  
 \cdot P(\alpha)
\end{equation}
where $n=N_{\textrm{pix}}$ and $\beta = (d-s)N^{-1}(d-s)$ is the
$\chi^2$. (Note that the $\chi^2$ is already calculated within
  the Gibbs sampler, as it is used to validate that the input noise
  maps and beams are within a correct range for each Gibbs
  iteration. Sampling from this distribution within the Gibbs sampler
  represent therefore a completely negligible extra computational
  cost.) For the Gaussian prior with unity mean and standard
deviation $\sigma_\alpha$, we find that
\begin{equation}
  P( \alpha \,|\, s, C_\ell, d ) \propto
  \frac{e^{-\frac{1}{2}(\frac{\beta}{\alpha} + \frac{ (\alpha-1)^2 }{
        \sigma_\alpha^2 })}}{\alpha^{n/2}}  
\label{eq:finally}
\end{equation}

For large degrees of freedom, $n$, the inverse gamma function
converges to a Gaussian distribution with mean $\mu = b/(k+1)$, where
we have defined $k = n_{\textrm{pix}}/2 -1$, and variance $\sigma^2 =
b^2/((k-1)(k-1)(k-2))$. A good approximation is therefore letting
$\alpha_{i+1}$ be drawn from a product of two Gaussian distributions,
which itself is a Gaussian, with mean and standard deviation
\begin{equation}
  \mu = \frac{\mu_1\sigma_2^2 + \mu_2\sigma_1^2}{\sigma_1^2 + \sigma_2^2}
\end{equation}
\begin{equation}
  \sigma = \frac{\sigma_1^2\sigma_2^2}{\sigma_1^2 + \sigma_2^2}.
\end{equation}

This sampling step has been implemented in ``Slave'', and we have
successfully tested it on simulated maps. With $N_{\textrm{side}}=512$ and
$l_{\textrm{max}}=1300$ and full sky coverage, we find $\alpha = 1.000 \pm
0.001$. Note that with such high resolution, the standard deviation on
$\alpha$ is extremely low, and any deviation from the exact $\alpha =
1.0$ will be detected. 

\section{Data}
\label{sec:data}

In this paper we analyze the 5-year WMAP data\citep{hinshaw:2009},
which is available from
LAMBDA\footnote{http://lambda.gsfc.nasa.gov}. We consider both the raw
sky maps for each differencing assembly (DA), and the corresponding
foreground-reduced maps \citep{gold:2009}. The nominal noise
amplitudes are taken from the original 5-year data release as
presented on LAMBDA. However, we note that after the initial
publication of this paper, these have been corrected, as there was a
discrepancy between the values presented on LAMBDA and those published
in the WMAP Five Year Explanatory Supplement \citep{limon:2009}. The
updated values agree very well with our results.

The foreground-reduced maps were produced from the raw maps by fitting
and subtracting three fixed templates to each case. One of these
templates was the (K-Ka) difference map, which mainly traces
synchrotron, free-free and, possibly, spinning dust, and the other two
were the H$\alpha$ template of \citet{finkbeiner:2003} and the FDS8
thermal dust template of \citet{finkbeiner:1999}. Here it is worth
noting that the (K-Ka) difference map was smoothed to $1^{\circ}$ FWHM
before fitting \citep{hinshaw:2007}; although this difference map is
intrinsically noisy, it does not have power beyond $\ell \sim 400$,
and, in effect, the raw and foreground-corrected maps are identical at
high $\ell$'s. This will be explicitly demonstrated later.

In the low signal-to-noise regime, the estimated CMB signal will
fluctuate greatly. In order to dampen the unruly behavior in this
regime, we have chosen to bin the power spectrum for high $\ell$s. The
$C_\ell$s are then generated from the binned signal power spectrum.

Unless explicitly noted, we apply the KQ85 WMAP sky cut
\citep{gold:2009}, which removes 18\% of the sky, including point
source cuts. We also take into account the circular-symmetric beam
profiles for each DA. Finally, the main analysis is carried out at a
HEALPix\footnote{http://healpix.jpl.nasa.gov} resolution of
$N_{\textrm{side}}=512$ and included harmonic space multipoles between
$\ell=2$ and 1300.  Including such high multipoles in the multipole
expansion is acceptable in this case for two reasons; first, the WMAP
beams fall off quickly in this regime, and the data becomes strongly
noise dominated. Second, we bin the angular power spectrum heavily at
high $\ell$'s.

\section{Results}
\label{sec:results}

\begin{deluxetable}{lccccc}
\tablewidth{0pt}
\tablecaption{Noise estimation results \label{tab:main}} 
\tablecomments{Summary of 5-year WMAP noise amplitudes. Second column: Noise amplitudes estimated directly from
  high-$\ell$ power spectra, quoted in terms of $\alpha-1$ in
  percent. Third column: Noise amplitudes estimated by Gibbs
  sampling. The uncertainty on each of these numbers is
  $0.1\%$. Fourth column: Noise RMS per observation as quoted by the
  WMAP team on LAMBDA. Fifth column: Noise RMS per observation
  estimated by Gibbs sampling. sixth column: Noise RMS per
  observation as quoted by the WMAP team in the Five Year Explanatory
  Supplement, page 65.}
\tablecolumns{6}
\tablehead{  &  \multicolumn{2}{c}{$\alpha-1$ (in \%)} & \multicolumn{2}{c}{$\sigma_{0}$ (mK)}\\
Band &  Direct & Bayesian & Nominal & Estimated & Updated}
\startdata
\cutinhead{Raw maps}
K1     & $1.90$  & $0.50$  & $1.436$ & $1.439$ & \\
Ka1    & $0.10$  & $0.10$  & $1.470$ & $1.471$ & \\
Q1  & $0.23$  & $0.15$     & $2.254$ & $2.256$ &\\ 
Q2  & $0.16 $ & $-0.10$    & $2.141$ & $2.140$ &\\ 
V1  & $0.08$  & $-0.08$    & $3.314$ & $3.313$ &\\ 
V2  & $-0.06$  & $-0.03$   & $2.953$ & $2.953$ &\\ 
W1  & $0.52 $ & $-0.37$    & $5.899$ & $5.889$ &\\ 
W2  & $0.23 $ & $-0.43$    & $6.565$ & $6.550$ &\\ 
W3  & $0.33 $ & $-0.30$    & $6.926$ & $6.916$ &\\ 
W4  & $-0.45 $ & $-0.85$   & $6.761$ & $6.732$ &\\

\cutinhead{Foreground-reduced sky maps}
Q1 &   $0.83$ & $0.51$  & $2.2449$ & $2.2542$ & $2.254$ \\ 
Q2 &   $0.87$ & $1.02$  & $2.1347$ & $2.1455$ & $2.141$ \\ 
V1 &   $0.71$ & $0.50$  & $3.3040$ & $3.3123$ & $3.314$ \\ 
V2 &   $0.63$ & $0.62$  & $2.9458$ & $2.9549$ & $2.953$ \\ 
W1 &   $1.06$ & $0.53$  & $5.8833$ & $5.8988$ & $5.899$ \\ 
W2 &   $1.27$ & $0.71$  & $6.5324$ & $6.5555$ & $6.565$ \\ 
W3 &   $1.54$ & $1.12$  & $6.8849$ & $6.9233$ & $6.926$ \\
W4 &   $0.04$ & $-0.20$ & $6.7441$ & $6.7373$ & $6.761$ 

\enddata
\end{deluxetable}

The main results from our analysis are given in the third column of
Table \ref{tab:main}, where the posterior mean of $\alpha$ is given
for each DA, both for raw and foreground-corrected maps. The numbers
are quoted in terms of $\alpha-1$ in percent, and the corresponding
posterior RMS is $0.1\%$.

For most of the raw DA sky maps, we find $\alpha = 1$ to within a few
sigma. The largest outlier is W4, with a negative amplitude of
-0.85\%. This DA is known to have the strongest correlated noise of
any WMAP DA. In general, we find that the noise levels of the raw sky
maps are in good agreement with the levels quoted by the WMAP team.

However, the situation is different for the foreground-corrected maps
relative to the original noise amplitudes presented on
  LAMBDA. Specifically, in general the amplitudes of these maps are
shifted high by $0.5-1.0\%$. The most extreme shift is seen for the W3
DA, with a posterior distribution centered on $\alpha=1.011\pm0.001$.

With such significant discrepancies between the predicted and observed
noise levels for the foreground-reduced sky maps, it should be
possible to observe an absolute shift in the high-$\ell$ power spectra
between simulated and the real sky maps. We therefore implemented a
simple and approximate method to estimate $\alpha$ directly, without
going through the Gibbs sampler: First, we calculate the angular power
spectrum for each DA to high $\ell$'s, where the beam has killed all
signal, and only noise is left. Any sky cut is ignored in this
step. We then compute the average spectrum amplitude at $\ell > 1300$,
and define this as our noise amplitude estimate. Next, we simulate an
ensemble of noise realizations from the RMS maps of each DA, and
repeat the above calculation for each of these, and compute the
corresponding average. The ratio between observed noise spectrum
amplitude and the simulated average is then an estimate of $\alpha$,
and can be compared to the values obtained from the Bayesian
analysis. Again, note that this is only a rough cross-check on
  the our results, and should not be considered a proper stand-alone
  result because the sky cut is completely ignored.  


The results from this exercise are tabulated in the second column of
Table \ref{tab:main}. In general, we see that these generally follow
the same trends as those from the Bayesian analysis, although with
systematically slightly higher amplitudes. This is likely due to
neglecting of the galaxy cut, as foregrounds may obviously add to the
total spectrum level. 



\begin{figure}
\mbox{\epsfig{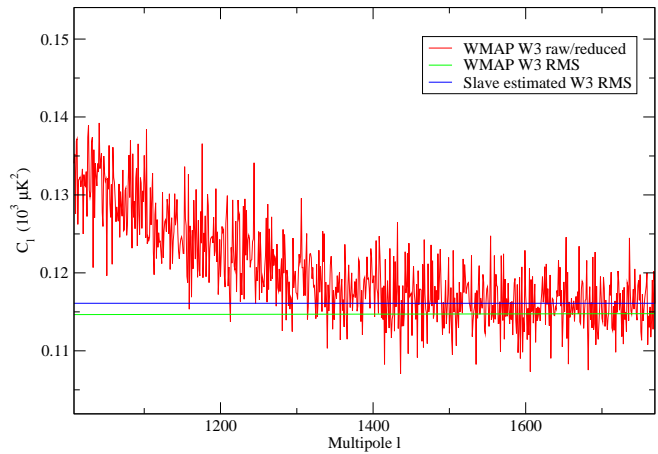}}
\caption{The raw/foreground reduced power spectrum (red) plotted with
  the WMAP noise (green) and the Slave-estimated noise (blue). Note
  how Slave estimates a different noise level (blue) than the input
  noise level (green) .} 
\label{fig:noisespectrum}
\end{figure}

In figure \ref{fig:noisespectrum} we plot the power spectrum of the
raw 5-year WMAP W3 sky map in $C_l$ units, together with the average
W3 RMS $\sigma_0$ value, as given by the WMAP team. We also plot the
estimated new W3 RMS, as estimated by Slave. Here it is clear that the
Slave estimate fits the power spectrum better. 

\begin{figure}
\mbox{\epsfig{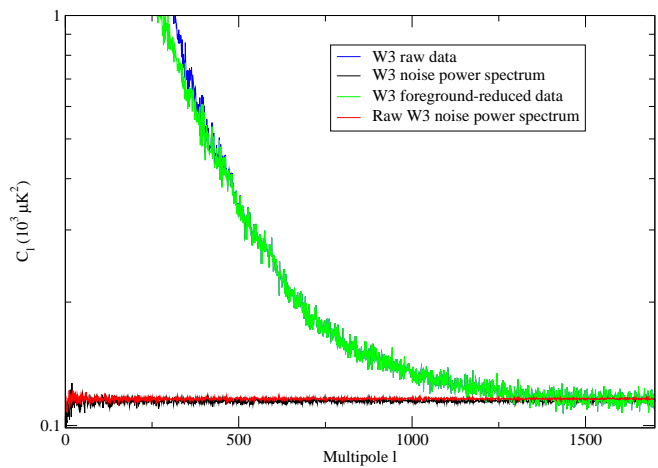}}
\caption{A noise power spectrum realization for W3 RMS (black) and raw
  W3 RMS (red). The data power spectrum can be
  seen entering at high $\ell$ (blue and green), where the raw and
  foreground-corrected data converge.}
\label{fig:noisecompare}
\end{figure}

Finally, in Figure \ref{fig:noisecompare} we plot the same
information, but to lower $\ell$'s and on a logarithmic scale. Notice
how the high-$\ell$ W3 foreground reduced maps power spectrum equals
the spectrum from the raw map. This implies that both the
foreground-reduced and raw data have the same high-$\ell$ noise
levels, and do not differ. We therefore conclude that the shift in the
noise amplitude is not due to any shift in the actual data, but rather
a difference in the WMAP5 noise models. 

In the fourth column of Table \ref{tab:main}, we tabulate for easy
reference the noise RMS values for a single observations listed on
LAMBDA. Here it is seen that the predicted noise levels for the
foreground corrected maps is indeed lower than those for the raw
maps. However, this does not seem to be the case, neither from looking
at the raw power spectra, nor from the results from our Bayesian
analysis. 

Again, following the first publication of this letter, it was
  found that the foreground-correct noise RMS values on the LAMBDA
  website were not properly updated with the 5-year WMAP
  results. Therefore, the noise levels used in this analysis are known
  to be incorrect, and the reason is fully understood. Thus, this
  demonstrates the new method works well as the estimated values are
  fully consistent with the corrected 5-year values, shown in the
  sixth column of Table \ref{tab:main}, column 6.

\section{Conclusion}
\label{sec:conclusion}

We have introduced a new method for estimating white noise levels in
CMB sky maps, using a Bayesian framework. We have then applied this
method to the 5-year WMAP data, and re-estimated the noise levels of
both the raw and foreground-reduced sky maps. In doing so, we found
that the predicted noise levels for the raw maps are in acceptable
agreement with the predictions, while the noise levels in the
foreground-reduced maps are $0.5 - 1.0\% $ higher than the estimate
initially provided by the WMAP team on LAMBDA. 

The explanation for this effect has after the publication of
  this paper been found by the WMAP team simply to be an error in the
  results provided on LAMBDA: The quoted values were derived from the
  3-year analysis instead of the 5-year analysis. However, the correct
values were used in their cosmological analysis for the 5-year data,
and no results are therefore compromised by this error.

Thus, the method has already been demonstrated to be both accurate and
useful on a practical example. Further, it carries virtually no extra
computational cost within a Gibbs sampler, since all required
quantities are already computed within this algorithm. We therefore
recommend this feature to be used as a standard part of the Gibbs
sampling machinery, since it both provides additional robustness
against noise mis-estimation, and also propagates the uncertainty in
these estimates into the final CMB power spectrum.

Finally, we note that these noise estimates are very robust with
respect to systematic issues such as foreground or CMB signal
estimation. The reason is that the sky maps are well oversampled; with
$\ell_{\textrm{max}}=1300$ there are $\sim1.7$ million modes in the
harmonic expansion, while at $N_{\textrm{side}}=512$ there are $\sim3$
million pixels. Therefore, there are essentially 1.3 million modes
available to estimate one single number, $\alpha$. If even greater
precision is desired, one could simply consider increasing the pixel
resolution one step further, which would leave the sky signal
unchanged, because it is bandwidth limited, whereas the total number
of modes in the map quadruples, thus decreasing the noise uncertainty.

\begin{acknowledgements}
  The authors wish to thank Joanna Dunkley for useful comments
    and for pointing out the issue of incorrect noise levels on
    LAMBDA. NEG and HKE acknowledge financial support from the
  Research Council of Norway.  The computations presented in this
  paper were carried out on Titan, a cluster owned and maintained by
  the University of Oslo and NOTUR. We acknowledge use of the
  HEALPix\footnote{http://healpix.jpl.nasa.gov} software
  \citep{gorski:2005} and analysis package for deriving the results in
  this paper. We acknowledge the use of the Legacy Archive for
  Microwave Background Data Analysis (LAMBDA). Support for LAMBDA is
  provided by the NASA Office of Space Science.
\end{acknowledgements}

\clearpage

\clearpage

\end{document}